\documentclass[aps,prl,twocolumn]{revtex4-1}
\usepackage{amsmath}
\usepackage{epsfig}
\usepackage{graphics}
\usepackage{verbatim}
\usepackage{color}
\newcommand{\beq}{\begin{equation}}
\newcommand{\eeq}{\end{equation}}
\newcommand{\bp}{{\bf p}}
\newcommand{\bq}{{\bf q}}
\newcommand{\bk}{{\bf k}}

\newcommand{\bP}{{\bf P}}
\newcommand{\fcal}{{\mathcal F}}

\usepackage{verbatim}

\begin{document}
\title{Itinerant Ferromagnetism in a polarized two-component Fermi gas}
\author{Pietro Massignan}
\affiliation{ICFO - The Institute of Photonic Sciences - 08860 Castelldefels, Barcelona, Spain}
\author{Zhenhua Yu}
\affiliation{Institute for Advanced Study, Tsinghua University, Beijing 100084, China}
\author{Georg M. Bruun}
\affiliation{Department of Physics and Astronomy, University of Aarhus, Ny Munkegade, DK-8000 Aarhus C, Denmark}

\date{\today}
\begin{abstract}
We analyze when a
 repulsively interacting two-component
  Fermi gas 
becomes thermodynamically unstable against phase separation. 
We focus on the strongly polarised limit where the free energy of the homogeneous mixture can be calculated accurately in terms of well-defined quasiparticles, the repulsive polarons. 
Phase diagrams as a function of polarisation, temperature, mass imbalance, and repulsive polaron energy, as well as scattering length and range parameter are provided.
We show that the lifetime of the repulsive polaron increases significantly with the interaction range and the mass of the minority atoms, raising the prospects of detecting the transition to the elusive itinerant ferromagnetic state with ultracold atoms.

\end{abstract}
\maketitle

A mixture of two  species of fermions with short range strongly repulsive interactions 
was predicted long ago by E.~Stoner to display a transition to an itinerant ferromagnetic state~\cite{Stoner1933}.
In analogy with immiscible classical fluids, the two components spatially separate to minimise their interaction energy.
This strong coupling effect is notoriously difficult to analyze, and the conditions for the existence of itinerant ferromagnetism are still controversial. 
For instance, one can rigorously exclude ferromagnetism in one dimension~\cite{Lieb1962}, whereas it has been shown to exist in specific types 
of Hubbard models~\cite{Nagaoka1966,Mielke1991a,Mielke1991b,Mielke1992,Tasaki1992,Tasaki1995}. 
First order transitions to a ferromagnetic state have been observed in  metallic systems as predicted by effective  theories~\cite{Belitz2012}. However, the inevitable presence
 of disorder and of intricate band structures in solid state systems imposes great difficulty when
  comparing microscopic theories with experiments, and it is still debated whether a homogenous electron system becomes ferromagnetic at all.

An exciting new setting to investigate itinerant ferromagnetism is provided by  quantum gases, which offer a precise determination 
of physical parameters and a great tunability \cite{Giorgini2008}. 
Conditions for itinerant ferromagnetism have been derived in the cold atom context using mean-field, diagrammatic, Jastrow-Slater, and Monte-Carlo calculations~\cite{Duine2005,LeBlanc2009,Conduit2009,Pilati2010,Cui2010,Massignan2011,
Chang2011,Heiselberg2011,Palestini2012,Cui2012,deSaavedra2012,Sodemann2012,He2012}. 
A recent experiment claimed the observation of itinerant ferromagnetism in a   $^6$Li mixture~\cite{Jo2009}. Current consensus however is that 
fast decay  to a lower lying attractive state precluded the realization of ferromagnetism in this experiment~\cite{Pekker2011,Pekker2011bis, Zhang2011,Massignan2011,Sanner2012,Sodemann2012}. Such decay is peculiar to quantum gases as the repulsive states obtained by Feshbach resonances are always accompanied by a number of attractive ones.

A major step in the quest to obtain long-lived repulsive Fermi gases is the production of a $^6$Li-$^{40}$K mixture around a moderately narrow Feshbach resonance, which  leads to a much smaller decay rate of the mixture~\cite{Kohstall2012}. 
This result raises the hope of observing itinerant ferromagnetism in quantum gases at Feshbach resonances which are not broad.

\begin{figure}
\includegraphics[width=0.8\columnwidth]{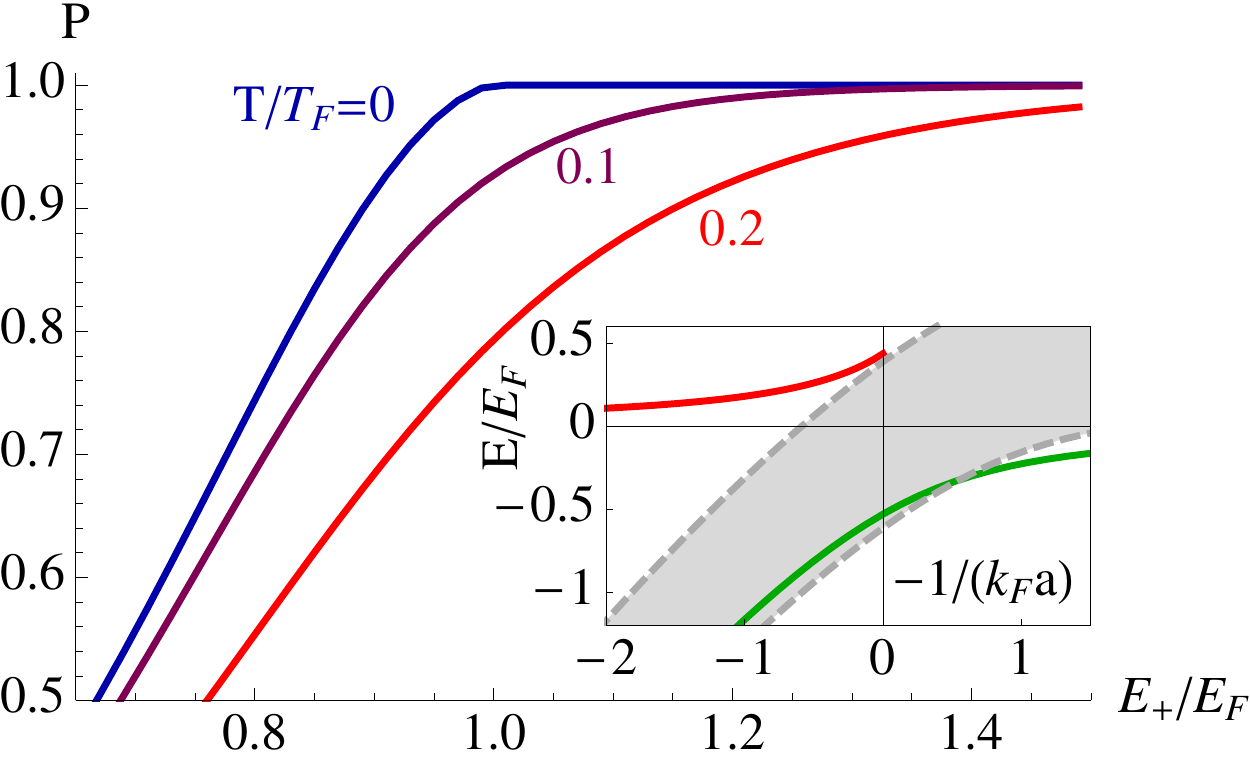} 
\includegraphics[width=0.8\columnwidth]{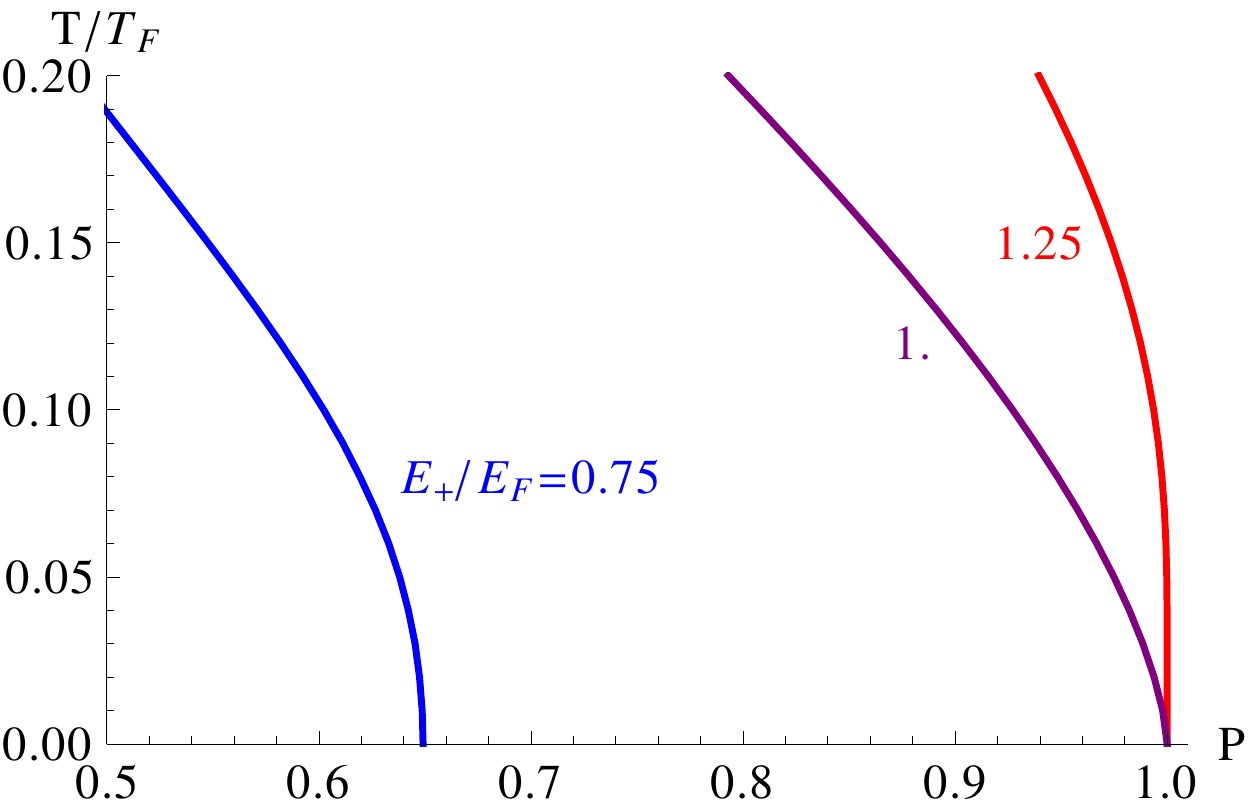}
\caption{Phase diagrams in terms of the polaron energy $E_+$, the polarization $P$, and the temperature $T$. The gas is mixed above the lines, and phase separated below. The inset shows the spectrum of a $^{40}$K impurity in a$^6$Li Fermi gas with $k_FR^*=1$, as in the experiment of Ref.\ \cite{Kohstall2012}: visible are the repulsive and attractive polarons (red and green lines), and the molecule-hole continuum (grey shaded region), arising from holes and molecules with momenta $0<q<k_F$.}
\label{fig:phaseDiagramVsEplus}
\end{figure}

In this letter, we examine the conditions for the existence of itinerant ferromagnetism for a two-component Fermi gas consisting of $N_1$ $1$-fermions with mass $m_1$ 
and $N_2$ $2$-fermions with mass $m_2$ in a volume $V$. 
The key point of our paper is that  we focus on the limit of strong polarisation taking $N_2\ll N_1$. 
An important result emerging from recent cold atom investigations is that it is possible to develop a quantitatively accurate description of the  properties of the gas in this limit, in terms of well-defined quasiparticles.
 The quasiparticles consist of 2-atoms dressed by multiple excitations of the bath of 1-atoms.
 The relevant quasiparticle spectrum as a function of the scattering length of the atom-atom interaction is shown in the inset of Fig.~\ref{fig:phaseDiagramVsEplus}. 
The spectrum contains a ``repulsive branch" at positive energy, populated by quasiparticles known as repulsive polarons. 
There is also an ``attractive" branch at negative energy and the corresponding quasiparticles are called attractive polarons. In addition, there is a  continuum 
of states (grey shaded region in the inset of Fig.\ \ref{fig:phaseDiagramVsEplus}a) corresponding to molecules and holes with momenta $0<q<k_F$ \cite{Kohstall2012} (or trimers plus two holes for light impurities \cite{Mathy2011}). 
Recent investigations showed that the polaron properties can be calculated  with surprising accuracy in terms of microscopic parameters
\cite{Combescot2007,Prokofev2008,Combescot2008,Chevy2010,Cui2010,Massignan2011,
Mathy2011,Massignan2012,Vlietinck2013}. 
It is the presence of a repulsive branch which opens the possibility to study  repulsively-interacting systems with cold gases.

We derive the free energy of the homogeneous mixture in terms of the repulsive polaron energy and analyze when the homogeneous mixture becomes unstable against phase separation.
Phase  diagrams are obtained as a function of polarisation, temperature, mass-imbalance and polaron energy as well as scattering length and range parameter. 
We show that the lifetime of the polarons increases significantly with the range parameter of the interaction.
This increase in the lifetime provides a much longer time-window for experiments,
suggesting that itinerant ferromagnetism might finally be
 unambiguously detected with cold atoms.

We take the interaction between the 1- and 2-atoms to be strong and repulsive, and the intra-species interaction to be negligible.
For a highly polarised mixture with $y=N_2/(N_1+N_2)\ll 1$, an accurate many-body theory describes the mixture as an ideal gas 
of $1$-atoms coexisting with an ideal gas of repulsive polarons~\cite{Pilati2010,Cui2010}.
In this limit, the free energy per particle reads
\begin{align}
\fcal_{\rm mix}=&(1-y)\fcal_{1}(n_1)+y \fcal_{2}(n_2)+y(1-y)^{2/3}E_{+}
\label{energyPPmix}
\end{align}
where $\fcal_\sigma(n_\sigma)=\mu_\sigma-k_BT {\rm Li}_{5/2}(-z_\sigma)/{\rm Li}_{3/2}(-z_\sigma)$ is the free energy per particle of an ideal gas 
of $\sigma$-atoms with mass $m_\sigma$ and density $n_\sigma=N_\sigma/V$ at temperature $T$, and ${\rm Li}_x(z)$ is the polylogarithm.
 The fugacity $z_\sigma=\exp(\mu_\sigma/k_BT)$, with $\mu_\sigma$ the chemical potential, is determined by $n_\sigma=-{\rm Li}_{3/2}(z_\sigma)/\lambda^3_\sigma$ where $\lambda_\sigma=(2\pi\hbar^2/k_BTm_\sigma)^{1/2}$ is the thermal de Broglie wavelength.
 $E_+$ is the energy of a zero-momentum repulsive polaron embedded in a Fermi sea of $1$-atoms with density $n=N/V$ and Fermi energy $E_{F}=\hbar^2k_F^2/2m_1$, with $k_F=(6\pi^2 n)^{1/3}$.
  The factor $(1-y)^{2/3}$ in Eq.\ (\ref{energyPPmix}) is a rescaling of the polaron energy taking into account the fact that the impurities are immersed in a Fermi sea with density $n_1=(1-y)n<n$.
In the weak coupling limit  $k_Fa\ll 1$, the polaron energy is given by  $E_+=2\pi a n_1/m_r$ with $m_r^{-1}=m_1^{-1}+m_2^{-1}$. The
 strong coupling expression for $E_+$ is given in the Supplemental Material.
 Since we focus on the low-$T$ regime, we have neglected the temperature dependence of the 
polaron energy $E_+$ which is approximated  by its $T=0$ value.
 The effective mass $m_{2,+}^*$ of the repulsive polaron is close to $m_2$~\cite{Massignan2011,Massignan2012}, and we have
taken $m_{2,+}^*=m_2$ for simplicity. Quantum Monte-Carlo (QMC) calculations~\cite{Pilati2010} indicate that the expression (\ref{energyPPmix}) is accurate for polarizations $P=(N_1-N_2)/(N_1+N_2)\gtrsim 0.5$ (i.e., $y<1/4$).

The thermodynamically unstable region of the homogeneous mixed phase can be determined by applying the usual Maxwell construction to the free energy $\fcal_{\rm mix}$.
Let us first focus on the case of equal masses $m_1=m_2$. The case of $m_1\neq m_2$ will be treated later. For equal masses, one has $\fcal_{\rm mix}(y)=\fcal_{\rm mix}(1-y)$. If $\fcal_{\rm mix}$ given by  (\ref{energyPPmix}), applicable in the regime $y\lesssim1/4$, exhibits a minimum at $y_0>0$, the homogeneous mixed phase become thermodynamically unstable when $y>y_0$.

In Fig.~\ref{fig:phaseDiagramVsEplus}, we plot the phase diagrams obtained from a Maxwell construction using Eq.\ (\ref{energyPPmix}) as a function of the polaron energy $E_+$, the polarisation $P$, and the temperature $T$.
We have only shown the diagram for $P\ge 1/2$, where our theory can be expected to be accurate.
For $T= 0$ and $P\to 1$, Fig.~\ref{fig:phaseDiagramVsEplus} shows
that the system phase separates when $E_+> E_{F}$. This reflects that the $2$-atoms cannot diffuse into a polaron state in the ideal gas of $1$-atoms if the polaron energy is higher than the Fermi energy.
 For smaller  $P$, phase separation occurs at a smaller polaron energy $E_+$ since the system can 
 separate into two partially polarised phases, which reduces the kinetic energy cost.
 Conversely, we see that phase separation is  suppressed at higher temperatures due to the entropy of mixing.
 Note that the phase diagram in Fig.~\ref{fig:phaseDiagramVsEplus} is generic since it is based only on the existence of well-defined and long-lived quasiparticles with energy $E_+$, an assumption verified experimentally in Ref.~\cite{Kohstall2012}.

We now turn to study the effects of unequal masses $m_1\neq m_2$. For simplicity, we consider
$T=0$ and  the limit $P\rightarrow 1$ ($y\to0$). A sufficient condition for the instability of the homogeneous mixed state can be found by comparing its energy with that of a fully phase separated state, where the 1-atoms occupy a volume $V_1$ and the impurities occupy a volume $V_2 = V-V_1$.
Equating the pressures $p_\sigma$ of the two fully separated phases, and using $p_\sigma\propto (N_\sigma/V_\sigma)^{5/3}/m_\sigma$ for an ideal Fermi gas,
  we find $N_1/V_1=n [y(m_1/m_2)^{3/5}+1-y]$ and $N_2/V_2=n [(1-y)(m_2/m_1)^{3/5}+y]$.
To first order in $y$, the energy per particle of the fully phase separated state becomes
\begin{equation}
\varepsilon_{\rm sep}= E_{F}\left[3/5 + y(m_1/m_2)^{3/5}-y\right],
\label{Phasesepmassimb}
\end{equation}
which  is accurate when $y\ll 3|(m_1/m_2)^{3/5}-1|^{-1}$.
To the same order in $y$, from Eq.~(\ref{energyPPmix}), the energy per particle of the homogeneous mixed state at $T=0$ is
\begin{align}
\varepsilon_{\rm mix}=E_{F}\left[3/5 + y(E_+/E_{F})-y\right].
\end{align} 
The homogeneous mixture is unstable when $\varepsilon_{\rm mix}>\varepsilon_{\rm sep}$ which gives~\cite{Massignan2011}
\beq
E_+> \left(\frac{m_1}{m_2}\right)^{3/5}E_{F}=\frac{\hbar^2(6\pi n)^{2/3}}{2m_2^{3/5}m_1^{2/5}}.
\label{IFMcondition}
\eeq
This result indicates that for fixed $E_+$, phase separation is favored for heavy atomic species due to the suppression of the Fermi pressures. This effect was also discussed in Refs.~\cite{vonKeyserlingk2011,Cui2012}.
An analysis of the ferromagnetic transition by the virial expansion, valid at high temperatures, is presented in the Supplemental Material.

In quantum gases, while the interaction potential is generally attractive, effective repulsive interactions can be obtained by preparing the $2$-atoms on the  repulsive branch 
of the Feshbach resonance depicted in the inset of Fig.~\ref{fig:phaseDiagramVsEplus}~\cite{Jo2009,Kohstall2012}. 
The effective inter-species interaction can be characterised by the $s$-wave scattering length $a$ and the range parameter $R^*=\hbar^2/(2m_ra_{\rm bg}\delta\mu\Delta B)>0$, where $a_{\rm bg}$ is the background scattering length, 
$\delta \mu$ is the differential magnetic moment, and $\Delta B$ the magnetic width of the Feshbach resonance~\cite{Petrov2004,Bruun2005}.
At the many-body level, a small/large  $k_FR^*$ corresponds to a wide/narrow  Feshbach resonance. 

To make a direct connection with experiments, it is useful to express $E_+$ in terms of the physical parameters $a$ and $R^*$.
A very attractive feature of the high polarization limit is that it is possible to calculate  $E_+$  precisely  even for strong 
 interactions.
The explicit expression of $E_+(k_Fa,k_FR^*,m_\downarrow/m_\uparrow)$, obtained through a many-body calculation including one-particle-hole excitations from the Fermi sea of the majority atoms, has been derived and discussed in detail in Refs.~\cite{Cui2010,Massignan2011,Kohstall2012, Massignan2012},  and is reported for completeness in the Supplemental Material.

  \begin{figure}
\includegraphics[width=\columnwidth]{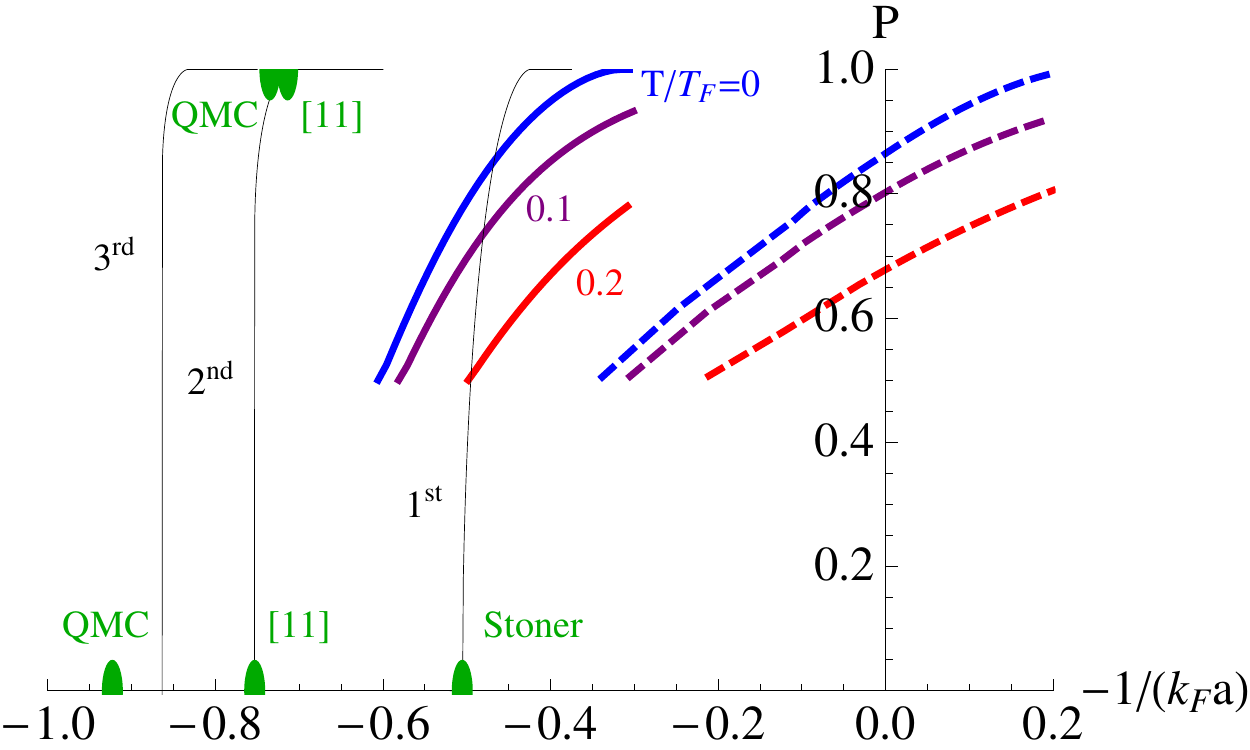}
\caption{Phase diagram for equal masses for $k_FR^*=0$ (thick solid lines), and $k_FR^*=1$ (thick dashed lines). The gas is mixed above the lines, and phase 
separated below. The thin lines are the phase boundaries obtained from the perturbative expression (\ref{Bishop}) retaining terms up to $1^{\rm st}$, $2^{\rm nd}$, and $3^{\rm rd}$ 
order. The green markers indicate the IFM transition at $T=R^*=0$ as found by Quantum Monte-Carlo calculations~\cite{Pilati2010}, second order diagrammatic theory~\cite{Duine2005},
 and Stoner~\cite{Stoner1933} for $P=0$ and $P=1$ respectively. 
}
\label{fig:phaseDiagramClassical_vs_minusOneOverkFa}
\end{figure}

We show the resulting phase diagram for equal masses in terms of the scattering parameters in Fig.~\ref{fig:phaseDiagramClassical_vs_minusOneOverkFa}
for two different values of the range: $k_FR^*=0$ describing a wide Feshbach resonance, and $k_FR^*=1$ similar to the moderately narrow Feshbach resonance of the experiment in Ref.~\cite{Kohstall2012}.
We observe that a non-zero effective range shifts the phase separated region toward the BCS-regime (the right side of the plot), consistent with a similar shift of the polaron energy reported in Ref.~\cite{Kohstall2012}.
As in Fig.~\ref{fig:phaseDiagramVsEplus}, we have drawn the phase boundary lines only within the regime of validity of the polaron theory, i.e.,~$P>1/2$.
With increasing $-(k_Fa)^{-1}$, we have furthermore terminated the lines where the polaron Ansatz fails due to decay as will be discussed in more detail below. 
For comparison, in Fig.\ \ref{fig:phaseDiagramClassical_vs_minusOneOverkFa}, we also mark the critical values for the ferromagnetic transition calculated by 
various zero-temperature theories. The perturbative results shown in the figure are obtained by inserting in Eq.~(\ref{energyPPmix}) the analytic expansion for the energy of a single impurity in a Fermi gas with hard-spheres interaction \cite{Bishop1973},
\beq
\frac{E_+}{E_F}=\frac{4 k_Fa}{3 \pi }
+\frac{2 (k_Fa)^2}{\pi ^2}
   +\left(\frac{4}{3}+\frac{2 \pi ^2}{45}\right)\frac{2  (k_Fa)^3}{\pi
   ^3}+\ldots
   \label{Bishop}
\eeq
 truncated respectively to $1^{\rm st}$, $2^{\rm nd}$, and $3^{\rm rd}$ order.  Note that these perturbative results should be taken with caution in the regime $|k_Fa|\sim1$. It is also known that the third order term in the expansion is non-universal.
It is interesting that Eq.~(\ref{energyPPmix}) combined with the single impurity expression (\ref{Bishop}) to $2^{\rm nd}$ order in $k_F a$ yields at $P=0$ essentially the same critical point as the $2^{\rm nd}$ order diagrammatic theory presented in Ref.~\cite{Duine2005}.

 Due to the symmetry of the phase diagram for equal masses, the  phase boundary  must cross the $P=0$ axis vertically.
 As can be seen from Fig.\ \ref{fig:phaseDiagramClassical_vs_minusOneOverkFa}, this  restricts
the range of possible extrapolations of our theory from its range of validity to smaller values of $|P|$. In particular, the extrapolation  to $P=0$ 
predicts a critical value for ferromagnetism inbetween the Monte-Carlo result and the mean-field  result. Our theory also predicts a larger critical value of $-1/k_F a$ than the QMC result for $P\rightarrow 1$. This is because the repulsive polaron energy obtained by our method is lower than the one found by QMC~\cite{Massignan2011,Pilati2010}, which is compatible with QMC being only an upper bound.
 Also, it could be due to higher order particle-hole processes omitted in our analysis. 

Our thermodynamical analysis above is based on the assumption that the repulsive polaron is a long-lived excitation.
However, the repulsive polaron state  inevitably experiences decay into the lower branches, i.e., the attractive polaron state or one with molecules or trimers formed (cf.~inset of Fig.\ \ref{fig:phaseDiagramVsEplus}a).
It is therefore important to address the question: does the repulsive polaron state have a 
sufficiently long lifetime in the vicinity of the phase transition 
 such that the experimental observation of the transition becomes possible?
This question has been considered for the balanced and polarized cases respectively in Refs.~\cite{Pekker2011,Pekker2011bis,Zhang2011,Sanner2012}, and \cite{Massignan2011,Sodemann2012}.
To take into account effects coming from a large $k_FR^*$ in the strongly-polarized regime, we calculate the decay rate $\Gamma_{\rm PP}$ of the repulsive polaron state 
due to the dominant two-body processes by using a diagrammatic method (see Suppl.\ Mat.\ for details)  which proved in very good agreement with the experimental data~\cite{Kohstall2012}.
We find that for fixed range parameter, the decay rate increases with increasing repulsion, and eventually the repulsive polaron state becomes ill-defined due to fast decay
as observed experimentally~\cite{Kohstall2012}. For this reason, at strong coupling we have terminated the lines in 
Fig.~\ref{fig:phaseDiagramClassical_vs_minusOneOverkFa} when $\Gamma_{\rm PP}/E_F>0.25$. However, most likely an even smaller decay is needed for the experimental observation of the ferromagnetic state. We now discuss how the latter may be achieved.

\begin{figure}
\includegraphics[width=0.8\columnwidth]{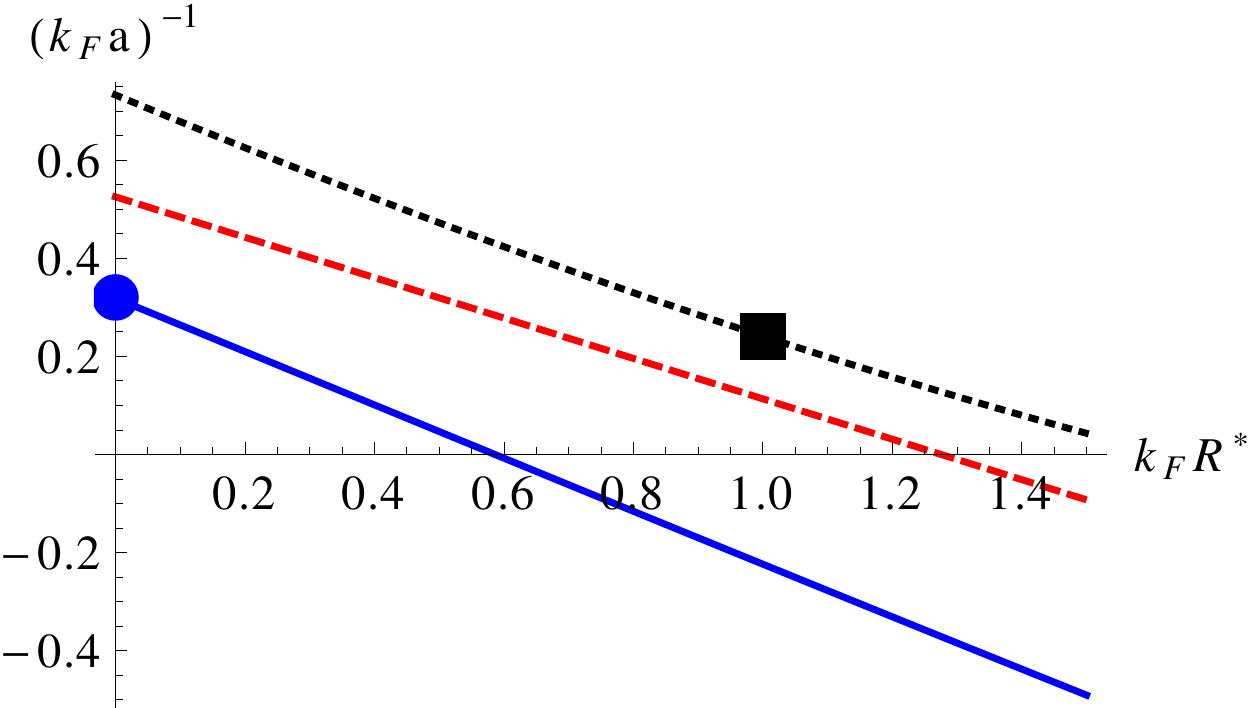}
\caption{Critical coupling strength  for phase separation at  $T=0$ and $P\rightarrow1$ as a function of $R^*$. Lines are for mass ratios $m_2/m_1=$1 (solid), 40/6(dotted), and 6/40 (dashed). The circle and the square indicate the values relevant for the experimental conditions of, respectively, Refs.~\cite{Jo2009} and \cite{Kohstall2012}.
}
\label{fig:IFMcondition}
\end{figure}

\begin{figure}
\includegraphics[width=0.8\columnwidth]{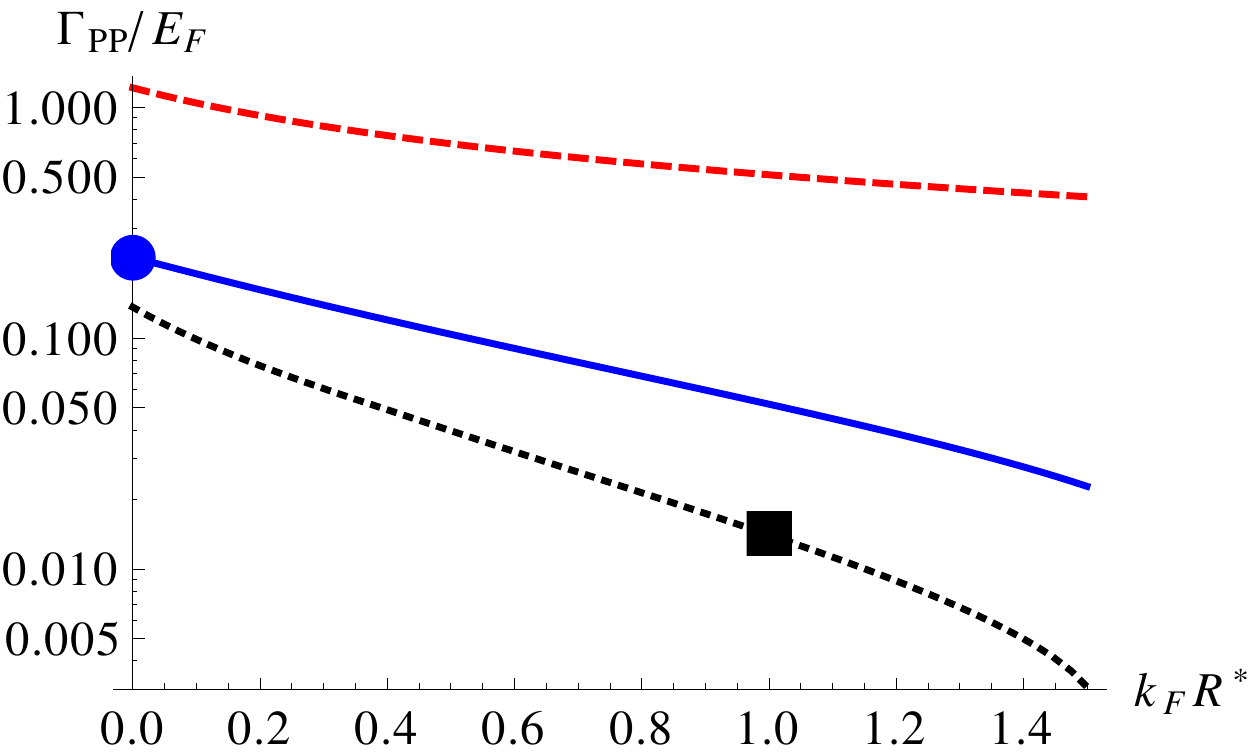}
\caption{Decay rate $\Gamma_{\rm PP}$ of repulsive polarons at the transition with $P\rightarrow1$ and $T=0$. Lines and symbols as in Fig.\ \ref{fig:IFMcondition}.}
\label{fig:decayRateAtTheTransition}
\end{figure}

We plot in Fig.\ \ref{fig:IFMcondition} the critical value of the interaction parameter $1/k_Fa$ for phase separation at $T=0$ and $P\rightarrow 1$ for different mass ratios 
obtained from (\ref{IFMcondition}), and in Fig.\ \ref{fig:decayRateAtTheTransition} the corresponding decay rate $\Gamma_{\rm PP}$ of the repulsive polaron at the critical coupling strength. 
Figure \ref{fig:decayRateAtTheTransition} shows an important result: 
 a resonance with  $k_F R^*\sim{\mathcal O}(1)$ gives rise to much longer lifetimes than a broad one with $k_F R^*\ll 1$.
 Furthermore, we find that a large mass ratio $m_2/m_1$ decreases the decay rate significantly compared to the 
case of equal masses. 
For instance, for a mixture of a small number of $^{40}$K atoms in a gas of $^6$Li atoms, the polaron lifetime increases by a factor $\sim 10$ at the critical coupling strength for phase separation when the range of the  atom-atom interaction is $k_FR^*=1$ instead of zero.
 By choosing appropriate mass ratios and using moderately narrow Feshbach resonances, one can suppress the decay rate of the mixed phase to $\sim E_F/100$.


 An unambiguous experimental signature of the itinerant ferromagnetism transition would be the observation of magnetic domains when the mixed gas is brought across
  the phase boundary. Domain formation was not detected in previous experiments~\cite{Sanner2012}, probably due to the fast decay with rate $\sim E_F/10$ suffered by 
  the atomic gas. The much slower decay of the repulsive polaron state predicted in this work when the range is significant can provide a ten-fold wider time window for 
  measuring the signatures of the transition, during which notable domains may form and be detected.  
  An estimation of the domain formation rate and size, though a worthy calculation, is out of the scope of the present paper. 
 The instability towards domain formation may also be detected by measuring
a sharp increase of 
 spin fluctuations. Experimentally, one can now measure
 spin fluctuations with $\mu$m-spatial resolution~\cite{Meineke2012} which indicates that this  method might be very effective.  
An alternative method for detection  is to prepare the gas directly in the phase separated state, and check its stability against mixing.

To conclude, we derived detailed phase diagrams of a two component Fermi gas in the limit of strong polarisation, where the effects of the interactions can be accurately described in terms of well-defined quasiparticles, the repulsive polarons. The phase diagrams were then expressed in terms of the scattering length and range parameter using a many-body theory known to be reliable in the strongly polarised limit. 
The ferromagnetic region was shown to move towards the BCS side with increasing range of the interaction. Finally, we found that a large range and impurity mass increase substantially the polaron lifetime, raising hopes for the observation of itinerant ferromagnetism in a quantum gas.

{\it Note added -- \/} Recently, we became aware of
related studies \cite{vonKeyserlingk2013} which have appeared after submission
of the present Letter.

{\it Acknowledgements -- \/} It is our pleasure to thank M.\ Zaccanti, T.-L.\ Ho and M. Lewenstein for insightful discussions.
This research has been funded through ERC Advanced Grant QUAGATUA, Spanish project TOQATA, Tsinghua University Initiative Scientific Research Program, NSFC under Grants 11104157 and 11204152, the Carlsberg Foundation and the ESF POLATOM network.

\bibliography{finiteTempImpurity}

\vspace{5cm}
\section{Supplemental material}
\subsection{Energy and decay rates of the repulsive polaron}
The quasiparticle properties of a polaron, i.e., a single impurity (2-atom) dressed by interactions with a Fermi sea of 1-atoms, can be derived from the knowledge of the impurity self-energy $\Sigma(\bp,E)$, which within the 1-particle-hole approximation and at zero temperature reads \cite{Chevy2010}
\beq
\Sigma(\bp,E)=\sum_{q<k_{F}}\mathcal{T}(\bp+\bq,E+\xi_{\bq,1}),
\label{selfEnergy}
\eeq
with $\xi_{\bk,\sigma}=k^2/(2m_\sigma)-\mu_\sigma$ the kinetic energy measured from the chemical potential (we have set $\hbar=1$).

The T-matrix for two-particle scattering reads~\cite{Bruun2005}
\begin{equation}
\mathcal{T}(\bP,\omega)=\frac{1}
{\left[
\frac{2\pi a_{\rm bg}}{m_r}\left(1-\frac{\Delta B} {B-B_0-E_{\rm CM}/\Delta\mu}\right)
 \right]^{-1} -\Pi(\bP,\omega)}
\label{T(mu,abg)}
\end{equation}
with $E_{\rm CM}=\omega-\bP^2/2M+\mu_1$ the energy in the center of mass reference frame of the scattering two particles, $M=m_1+m_2$, and the renormalized pair propagator
\beq
\Pi(\bP,\omega)=\sum_{\bk}\left[\frac{\theta(k-k_F)}{\omega+i0^{+}-\xi_{\bk,1}-\frac{(\bP+\bk)^2}{2m_2}}+\frac{2m_r}{k^2}\right].
\label{pairPropagator}
\eeq

The polaron energies at zero momentum are determined by the equation $\omega={\rm Re}[\Sigma(0,\omega)]$, which generally has a negative solution $E_-$, corresponding to the attractive polaron, and a positive solution $E_+$, corresponding to the repulsive one. The corresponding residues are given by $Z_\pm=\{1-{\rm Re}[\partial_\omega\Sigma(0,\omega)]_{\omega=E_{\pm}}\}^{-1}$.

In the leading decay process, a repulsive polaron decays into an attractive polaron by scattering a majority particle out of the Fermi sea~\cite{Massignan2011,Kohstall2012}.
To describe this 2-body decay channel, we use the fact that the impurity Green's function can be approximated as
\beq
 G^{\pm}(\bp,\omega)\sim\frac{Z_\pm}{\omega-E_\pm-p^2/2m_{2,\pm}^{*}}
\label{polaronsPoleExpansion}
\eeq
close to its poles $E_{\pm}$, with $m_{2,\pm}^{*}$ the effective mass of the polarons.
Replacing the bare impurity Green's function with $G^{-}$ inside the pair propagator Eq.\ (\ref{pairPropagator}),
\beq
\tilde\Pi(\bP,\omega)=Z_-\sum_{\bk}\left[\frac{\theta(k-k_F)}{\omega+i0^{+}-E_--\xi_{\bk,1}-\frac{(\bP+\bk)^2}{2m^*_{2,-}}}+\frac{2m_r}{k^2}\right],
\label{pairPropagatorDressed}
\eeq
and using $\tilde\Pi$ in Eqs.~(\ref{selfEnergy}) and (\ref{T(mu,abg)}), the decay rate of repulsive polarons to attractive ones is given by
\begin{equation}
\Gamma_{PP}=-2Z_+\mathrm{Im}[\tilde\Sigma(0,E_+)].
\label{2bodyDecay}
\end{equation}
For simplicity, in the calculation we have approximated $m^*_{2,-}$ with the bare value $m_2$.

\subsection{Virial expansion}
In the high temperature regime,
we examine the itinerant ferromagnetic transition by the virial expansion. 
To second order, this expansion gives 
for the partition function of the  two-component Fermi gas 
\begin{equation}
\mathcal Z=1+V\sum_{\sigma=1,2}\frac{z_\sigma}{\lambda_\sigma^3}+\frac{V}{2} \left(\sum_{\sigma=1,2}\frac{z_\sigma}{\lambda_\sigma^3}\right)^2+\frac{V z_1z_2 b_2}{\lambda_M^3},
\label{par}
\end{equation}
with $\lambda_M=(2\pi\hbar^2/k_BTM)^{1/2}$. When only the $s$-wave ``upper branch" of excitations is taken into account,
the second virial coefficient $b_2$ is \cite{Huang1987}
\begin{equation}
b_2=\frac{1}{m_rT}\int_0^\infty \frac{dk}\pi k \delta_s(k) e^{-k^2/2m_rT}.
\label{b2narrow}
\end{equation}
The $s$-wave phase shift is given by $\cot\delta_s(k)=-1/ka-R^*k$. At a broad resonance one has $b_2=e^{(\lambda_r/2\sqrt\pi a)^2}\left[1+{\rm erf}(\lambda_r/2\sqrt{\pi} a)-2\theta(a)\right]/2$, whose minimum value is $-1/2$ at $1/a=0^+$. 
For a very narrow resonance, we find $b_2\sim-\lambda_r/2\pi R^*$ at the unitary point. Here $\lambda_r=(2\pi\hbar^2/2k_BTm_r)^{1/2}$.

The free energy per particle of the mixture $\fcal_{\rm mix}=-k_BT\ln\mathcal Z/ N+(1-y)\mu_1 +y\mu_2$ is
\begin{align}
\fcal_{\rm mix}&=-k_BT\left\{1-(1-y)\ln[(1-y)n\lambda_1^3]-y\ln(yn\lambda_2^3)\right.\nonumber\\
&\left.+y(1-y)n\lambda_r^3 b_2\right\}, \label{FVirial}
\end{align}
where we have used $z_\sigma=\lambda_\sigma^3(n_\sigma-{\lambda_r^3}n_1 n_2 b_2)$ obtained from (\ref{par}).
The phase diagram will have a phase separated region when 
\begin{align}
\partial^2_y \fcal_{\rm mix}/k_BT=(1-y)^{-1}+y^{-1}+2 n\lambda_r^3 b_2\leq 0,
\label{instability_condition_from_2nd_order_virial}
\end{align}
for some values of $y$, which is possible only if $b_2<0$. For equal masses, the region of phase separation derived from the Maxwell construction is bounded by the 
concentrations which minimise $\fcal_{\rm mix}$, i.e.,
\begin{align}
\partial_y \fcal_{\rm mix}/k_BT=\ln[y/(1-y)]+(2y-1) n\lambda_r^3 b_2= 0.
\label{IFM_condition_from_2nd_order_virial}
\end{align}

\begin{figure}
\includegraphics[width=0.8\columnwidth]{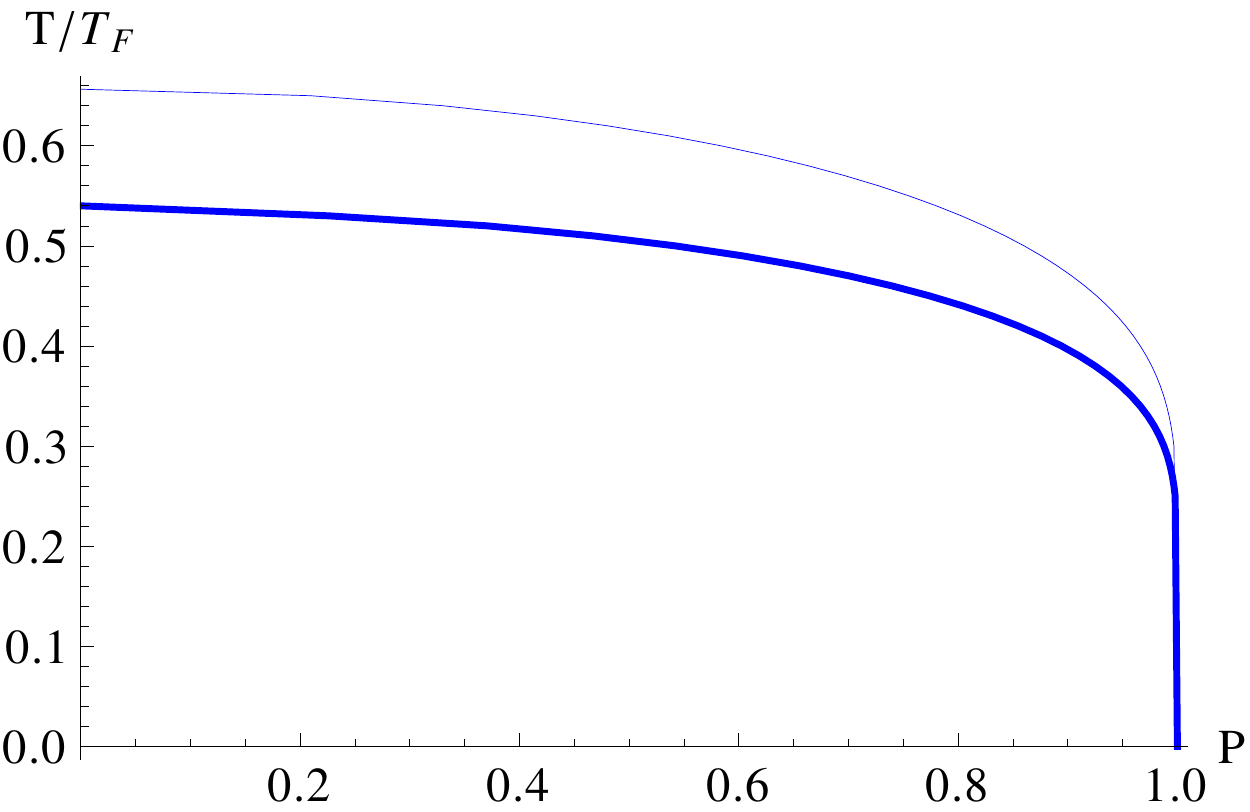}
\caption{Phase diagram at the unitarity point $(k_Fa)^{-1}=0^+$ for equal masses as obtained from the virial expansion to second order: $k_F R^*=0$ (thin line) 
and $k_F R^*=1$ (thick). The gas is mixed above the lines, and phase separated below.}
\label{fig:IFM_condition_from_2nd_order_virial}
\end{figure}

The phase diagram obtained from (\ref{IFM_condition_from_2nd_order_virial}) is plotted in Fig.~\ref{fig:IFM_condition_from_2nd_order_virial}
for $(k_Fa)^{-1}=0^+$ and $m_1=m_2$.
Since there is no kinetic energy cost of phase separation for high temperatures, this phase diagram is a result of the competition between the repulsive interaction energy and the entropy of mixing.
 The predicted critical temperature for phase separation unfortunately is too low for the virial expansion to be reliable.
Indeed, the largest transition temperature is $T_c^{\rm max}\approx0.66 T_F$ obtained at unitarity for a broad resonance where $b_2=-1/2$. 
The calculation nevertheless indicates that a large range decreases the critical temperature for phase separation, consistently with the low temperature polaron calculation.

   
\end{document}